\documentstyle[12pt]{ioplppt}

\begin{document}
\newcommand{\HF}{{\scriptsize HF}}
\newcommand{\FM}{{\scriptsize FM}}
\newcommand{\SSDW}{{\scriptsize SSDW}}
\newcommand{\DSDW}{{\scriptsize DSDW}}
\newcommand{\AFM}{{\scriptsize AFM}}
\newcommand{\FIELD}{\{\bDelta_{i},w_{i}\}} \newcommand{\kF}{k_{\rm F}}
\jl{3} \title{Spin density wave selection in the one-dimensional
Hubbard model}[Spin density waves in the 1D Hubbard model]

\author{J H Samson}

\address{Department of Physics, Loughborough University of
Technology, Loughborough, Leics LE11 3TU, UK
(Electronic address: j.h.samson@lut.ac.uk)}

\begin{abstract}
The Hartree-Fock ground state phase diagram of the one-dimensional
Hubbard model is calculated, constrained to {\it uniform} phases,
which have no charge density modulation.  The allowed solutions are
{\it saturated ferromagnetism} (\FM), a {\it spiral spin density wave}
(\SSDW) and a {\it double spin density wave} (\DSDW).  The \DSDW\
phase comprises two canted interpenetrating antiferromagnetic
sublattices.  \FM\ occurs for small filling, \SSDW\ in most of the
remainder of the phase diagram, and \DSDW\ in a narrow tongue near
quarter (and three-quarter) filling.  Itinerant electrons lift the
degeneracy with respect to canting angle in the \DSDW. The
Hartree-Fock states are metallic except at multiples of a quarter
filling.  Near half filling the uniform \SSDW\ phase is {\it unstable}
against phase separation into a half-filled antiferromagnetic phase
and a hole-rich \SSDW\ phase.  The dependence of the ground state wave
number on chemical potential is conjectured to be a staircase.
Comparison is made with higher dimensional Hubbard models and the
$J_{1}-J_{2}$ Heisenberg model.
\end{abstract}


cond-mat/9511116

To appear \JPCM

\maketitle

\section{Introduction}
The Hubbard Hamiltonian, originally proposed as a model of itinerant
magnets, has lately gained new interest as a possible Hamiltonian
for the cuprate superconductors.  The Hamiltonian is
\begin{equation}
	H=H_{0}+U\sum_{i}n_{i\uparrow}n_{i\downarrow}
	\label{Hubbard}
\end{equation}
where
\begin{equation}
	H_{0}=-\sum_{ijs}t_{ij}c_{is}^{\dag}c_{js}.
	\label{H0}
\end{equation}
Here $i$ and $j$ are site indices, $s=\uparrow,\downarrow$ is a spin
index and $U$ is the on-site Coulomb repulsion.  We shall be
considering the one-dimensional model with nearest-neighbour hopping,
$t_{ij}=t$ for $|i-j|=1$ and $t=0$ otherwise, and arbitrary band
filling $(0<n<2)$.  One of the few exact results is the Bethe Ansatz
ground state for this case (Lieb and Wu 1968).  An approximate
solution of the same model, especially one that breaks symmetries
present in the exact solution, therefore requires some justification.

The aim of the present work is to investigate spin structures in the
Hubbard model.  Analogous questions arise in frustrated Heisenberg
antiferromagnets, as reviewed by Schulz \etal (1994).  The simplest
example is the one-dimensional Hamiltonian
\begin{equation}
	H=J_{1}\sum_{i}{\bf S}_{i}\cdot{\bf S}_{i+1}+
	J_{2}\sum_{i}{\bf S}_{i}\cdot{\bf S}_{i+2}
	\label{MG}
\end{equation}
with $J_{1}, J_{2}$ positive (Majumdar and Ghosh 1969).  The classical
ground state is a spiral; in the limit $J_{1}\ll J_{2}$, the even and
odd sites decouple and the ground state becomes degenerate.  In the
spin $1/2$ case the peak in the structure factor moves from the
N\'{e}el value $\pi$ to $\pi/2$ with increasing $J_{2}$, as in the
classical case (Tonegawa and Harada 1987, Zeng and Parkinson 1995).
Unlike in the classical case, dimerization (local singlet formation)
occurs for $J_{2}/J_{1}$ of the order unity or larger (Haldane 1982).
The dimerization is not visible in two-spin correlation functions, but
is seen in singlet-singlet correlations of the form $\langle({\bf
S}_{0}\cdot{\bf S}_{1})({\bf S}_{i}\cdot{\bf S}_{i+1})\rangle$.

In certain situations the classical Heisenberg model possesses a line
of degenerate ground states (in addition to the global rotational
symmetry).  The standard example in two dimensions is a square lattice
antiferromagnet with nearest and next-nearest neighbour exchange
interactions, with $J_{1}<2J_{2}$.  The classical ground state here
comprises two interpenetrating antiferromagnetic sublattices, with no
coupling between the N\'{e}el vectors.  The three-dimensional face
centred cubic antiferromagnet is similarly frustrated, with collinear
spin density waves degenerate with a continuum of non-collinear double
and triple spin density waves (see e.g.  Long and Yeung 1986).  These
degeneracies could be lifted ``by hand'' by adding various terms to
the classical Hamiltonian, for example anisotropy or biquadratic
interactions.  Such terms are not needed; ``ordering by disorder''
stabilizes the collinear phase, the disorder being either quantum
(Shender 1982) or thermal (Henley 1987) in nature.  On the other hand,
Long (1989) and Henley (1989) have shown that non-magnetic impurities
favour islands of the non-collinear phase localized about the
impurity.  Both effects are due to the large transverse susceptibility
of each antiferromagnetic sublattice to fluctuations in the other
sublattice.  We will see here how an itinerant model provides another
mechanism for state selection.

Similar spin structures arise in the Hubbard model; indeed, this
reduces to the Heisenberg antiferromagnet in the half-filled
two-sublattice large-$U$ limit.  We shall be investigating
Hartree-Fock (\HF) solutions of the one-dimensional model.  The
paramagnetic \HF\ state is always unstable for $U>0$ in one dimension.
Perfect nesting ensures that the band susceptibility diverges at
$q=2k_{\rm F}$.  The system is therefore unstable towards a spin
density wave at this wave vector for infinitesimal $U$.  For finite
$U$ nonlinear effects can lead both to a shift in this wave vector and
a distortion.  We recognise that mean field theories in general, and
\HF\ calculations in particular, overestimate the tendency towards
magnetic order (which the Mermin-Wagner theorem forbids in one
dimension).  However, the ordering may survive in the form of
short-range correlations.

Most recent \HF\ studies of the Hubbard model relate to the
two-dimensional case, with reference to copper-oxygen planes in
cuprate superconductors.  An exhaustive search of solutions of the
\HF\ equations is impractical, with global minima hard to obtain and
in any case dependent on the boundary conditions.  Some authors
restrict consideration to collinear magnetization, following Machida
and Fujita (1984).  These found an exact solution for a
one-dimensional model with linearized dispersion: a {\it soliton
lattice} with a snoidal spin density wave.  Most studies find coplanar
spin textures in the two-dimensional Hubbard model; however, Chubukov
and Musaelian (1995) find evidence for non-coplanar textures.
Verg\'{e}s \etal (1991) identify a large number of stable or
metastable configurations.  {\it Spiral spin density waves} (\SSDW),
with wave vectors varying continuously with $U$ and $n$ (Dzierzawa
1992), become unstable near half-filling.  Here the \HF\ ground state
appears to be a collinear soliton lattice in which the holes are
localized on walls between N\'{e}el-ordered domains (Fujita \etal
1991, Ichimura \etal 1992).  The same conclusion follows from a
fourth-order Landau expansion, valid for weak coupling, with
coefficients determined from the electronic structure (Schulz 1990).
Nearest-neighbour Coulomb repulsion tends to stabilize the \SSDW\
phase against this hole clustering (Hu \etal 1994).  One may identify
various competing processes: firstly, a collinear spin density wave
will present the electrons with a spatially varying potential, which
may open a gap at the Fermi energy and stabilize the collinear wave.
On the other hand, large $U$ might favour the more uniform charge
density in a metallic \SSDW. Competing ordered states, such as charge
density waves and superconductivity, should also be taken into
consideration (Bach \etal 1994); however, for positive $U$ in the
absence of other interactions the \SSDW\ is favoured (Eriksson \etal
1995).  Some of the above features can already be seen in the
one-dimensional case discussed in this work.

Early interest of course concentrated on the three-dimensional model
(Penn 1966).  There have been studies of spin structures in itinerant
antiferromagnets, notably face-centred cubic $\gamma$-Mn, where single
and multiple spin density waves are degenerate ground states of the
classical Heisenberg model.  Hirai and Jo (1985) show how fourth order
terms in transfer integrals lift this degeneracy.  However, spin
density functional calculations reveal a very small energy difference
between these structures (Crockford \etal 1991).  Here \HF\ studies of
the one-dimensional Hubbard model can illustrate the origin of the
energy difference.

Bach \etal (1994) have studied a generalized unrestricted \HF\ theory,
in which the quadratic \HF\ Hamiltonian allows particle-nonconserving
terms and therefore treats magnetic and BCS states on an equal
footing.  They report a number of theorems concerning the ground state
symmetry, although not much is known rigorously about the finite-$U$
Hubbard model away from half filling.

The next section covers the solution of the \HF\ equations.  The
solutions are restricted to {\it uniform} states, in which the local
density of states is uniform up to spin rotation.  Such states fall in
two two-parameter families: {\it spiral spin density waves} (\SSDW)
and {\it double spin density waves} (\DSDW).  Each is parameterized by
a field amplitude and the nearest-neighbour angle.  The band structure
is calculated for both families, and the energy is minimized with
respect to the parameters.  Section \ref{pd} presents the resulting
phase diagram: the \DSDW\  is the ground state only in a narrow
region near quarter (and three-quarter) filling.  Near half filling,
the homogeneous \SSDW\ is unstable towards phase separation into a
half-filled antiferromagnetic domain and a hole rich (or electron
rich) domain.  Finally, section \ref{disc} discusses the physical
significance of the \DSDW\ and the stability of the states found.

A brief report of some aspects of this work has recently appeared
(Samson 1995).

\section{Computational method \label{calc}}
\subsection{The Hartree-Fock approximation}
The {\it unrestricted} \HF\ approximation minimizes
$\langle\FIELD|H|\FIELD\rangle$, the expectation value of the Hubbard
Hamiltonian \eref{Hubbard} in the space of Slater determinants
$|\FIELD\rangle$.  These states are ground states of the
non-interacting many-electron system in a spin- and site-dependent
Hamiltonian
\begin{equation}
	H_{\rm HF}(\FIELD)=H_{0}+\sum_{i}(-\bDelta_{i}\cdot
	{\bf {S}}_{i}+w_{i}\cdot n_{i}),
	\label{HHF}
\end{equation}
with
\begin{eqnarray}
	n_{i} & = & \sum_{s}c_{is}^{\dag}c_{is}  \\
	{\bf {S}}_{i} & = &
	\frac{1}{2}\sum_{st}c_{is}^{\dag}{\bf{\sigma}}_{st}c_{it}.
\end{eqnarray}
We take the number of sites $N_{\rm a}\rightarrow\infty$ and work in
the canonical ensemble, with a fixed number $n$ of electrons per site.
The general problem is a minimization of a function with a large
number of local minima and saddle-points, dependent on boundary
conditions and differing little in energy, in a $4N_{\rm
a}$-dimensional space.  We shall therefore be far less ambitious and
restrict consideration to {\it uniform} states, defined as those with
a site-independent spin-projected local density of states, referred to
the local spin quantization direction.  Defining the tight-binding
Green function in the usual way as
\begin{equation}
	[G_{ij}]_{st}(E)=\langle is|(E-H_{\rm HF})^{-1}|jt \rangle
	\label{Gij}
\end{equation}
where $|is\rangle$ is a one-electron Wannier orbital, we
require the local tight-binding Green function to be of the form
\begin{equation}
	G_{ii}(E)=a(E)+b(E){\bf e}_{i}\cdot{\bf \sigma}.
	\label{Gii}
\end{equation}
The only spatial dependence
allowed is in the local magnetization directions, given by the unit
vectors ${\bf e}_{i}$.  All atoms are
equivalent and there is no charge density modulation.  The
fields are then uniform in magnitude:
\begin{equation}
	w_{i}=\frac{1}{2}Un\ \ {\rm and}\ \ \bDelta_{i}=\Delta{\bf e}_{i}.
	\label{constraints}
\end{equation}
The energy then becomes

\begin{equation}
	E_{\rm HF}=\min_{\FIELD}V_{\rm HF}(\{\bDelta_{i},w_{i}\})
	\label{EHF}
\end{equation}
where the minimization is subject to the constraints
\eref{constraints} and
\begin{equation}
     V_{\rm HF}(\FIELD)=\frac{1}{N_{\rm a}}
     \left\langle\{\bDelta_{i},w_{i}\}\left|H_{\rm HF}\right|
     \{\bDelta_{i},w_{i}\}\right\rangle
	+\frac{\Delta^{2}}{4U}+\frac{1}{4}Un^{2}.
	\label{VHF}
\end{equation}
This is equivalent to the self-consistency condition
\begin{equation}
	\bDelta_{i}=2U\left\langle\FIELD\left|{\bf
	{S}}_{i}\right|\FIELD\right\rangle
	\label{sc}
\end{equation}
(so that in the \HF\ solution the field is parallel to the magnetization).

\subsection{Spin density waves}
The uniformity condition \eref{Gii} implies restrictions on the allowed
directions, as we see from the expansion
\begin{equation}
	G_{ii}(E)=g_{ii}(E)-\frac{1}{2}\Delta\sum_{j}g_{ij}(E){\bf
	e}_{j}\cdot{\bf
	\sigma}g_{ji}(E)+\ldots
	\label{Dyson}
\end{equation}
where $g_{ij}(E)$ is the tight-binding Green function of the band
Hamiltonian $H_{0}$ \eref{H0}. Consistency between equations
\eref{Dyson} and \eref{Gii} for all energies requires the conditions
\begin{equation}
	({\bf e}_{i-k}+{\bf e}_{i+k})\parallel{\bf e}_{i}\ \ \ \forall i,k.
	\label{parallel}
\end{equation}
The configurations are therefore coplanar, which implies vanishing
torque on the local moments (Small and Heine 1984).  (Non-coplanar
configurations are allowed in two or more spatial directions.) This
allows only two classes of configuration, as shown in \fref{configs}:
\begin{enumerate}
	\item  Spiral spin density wave (\SSDW)
	\begin{equation}
		{\bf e}_{i}=(\sin Qi,0,\cos Qi).
		\label{SSDW}
	\end{equation}

	\item  Double spin density wave (\DSDW)
	\begin{eqnarray}
		{\bf e}_{2k} & = & (-1)^{k}(0,0,1) \nonumber  \\
		{\bf e}_{2k+1} & = & (-1)^{k}(\sin\theta,0,\cos\theta).
		\label{DSDW}
	\end{eqnarray}
\end{enumerate}
(We take the lattice parameter to be 1, and the spins to be in the
$xz$ plane.) The \SSDW\ is specified by two parameters, the field
amplitude $\Delta$ and the pitch angle $Q$, varying from 0 (the
ferromagnetic phase $\uparrow\uparrow\uparrow\uparrow\ldots$) to $\pi$
(the antiferromagnetic phase
$\uparrow\downarrow\uparrow\downarrow\ldots$).  The \DSDW\ is
similarly specified by the amplitude $\Delta$ and an angle $\theta$.
It may be thought of as two interpenetrating antiferromagnetic
sublattices with the staggered magnetization vectors canted at an
angle $\theta$, varying between 0 (the collinear or dimerized
configuration $\uparrow\uparrow\downarrow\downarrow\ldots$) and
$\pi/2$, which coincides with the $Q=\pi/2$ \SSDW\
$\uparrow\rightarrow\downarrow\leftarrow\ldots$.  In the classical
Heisenberg model the molecular field of one sublattice on the other
vanishes in this phase, giving a two-dimensional manifold of
degeneracies; any translationally invariant two-spin correlation
function is independent of $\theta$.  The
\DSDW\ is also a superposition of two \SSDW\ states at $Q=\pm\pi/2$.

\subsection{Band structures}
This {\it restricted} \HF\ system is invariant under a subgroup of the
symmetry group of the Hubbard Hamiltonian.  Although translational and
spin rotational symmetry are both broken, the Hamiltonian remains
invariant under a simultaneous translation and spin rotation and
Bloch's theorem still applies.  We defer discussion of the
stability of such configurations against further symmetry breaking to
section \ref{stab}.

We calculate band structures of the \HF\ Hamiltonian \eref{HHF} in the
standard way.  Following Korenman \etal (1977), we transform the spin
quantization axis from the $z$ axis to the local axis
${\bf e}_{i}=(\sin\theta_{i},0,\cos\theta_{i})$:
\begin{equation}
	c_{is}=\sum_{t}U_{st}(\theta_{i})d_{it}
	\label{transformation}
\end{equation}
where
\begin{equation}
	U(\theta)=\left(
	\begin{array}{cc}
		\cos\theta/2 & -\sin\theta/2  \\
		\sin\theta/2 & \cos\theta/2
	\end{array}
	\right).
	\label{U}
\end{equation}
The transfer integrals $t_{ij}$ couple the $\d_{\uparrow}^{\dag}$ and
$\d_{\downarrow}^{\dag}$ bands, and the band structures reduce to the
solution of second- and fourth-order secular equations for the \SSDW\
and \DSDW\ respectively.  A little algebra leads to the two bands
\begin{equation}
	E_{\rm SSDW}(k)=-2t \cos(Q/2) \cos k \pm
	\sqrt{\Delta^{2}/4+4t^{2}\sin^{2}(Q/2)\sin^{2}k}
	\label{EkS}
\end{equation}
(with Brillouin zone $-\pi<k\le\pi$) in the \SSDW\ phase and the four bands
\begin{equation}
\fl	E_{\rm DSDW}(k)=\pm \sqrt{\Delta^{2}/4+2t^{2}
    \pm \sqrt{\Delta^{2}t^{2}(1+\sin\theta
    \cos 2k ) + 4t^{4}\sin^{2}2k}}
	\label{EkD}
\end{equation}
(with Brillouin zone $-\pi/2<k\le\pi/2$) in the \DSDW\ phase.

\Fref{bands} shows sample band structures, illustrating how
dimerization opens gaps in the folded \SSDW\ bands.  The Fermi surface
follows from the condition that the length of $k$ space occupied is
$2\pi n$.  For both families the Fermi surface always consists of
either 0, 2 or 4 points.

\subsection{Energy computation}
The energy of these configurations, following equation \eref{VHF}, is
\begin{equation}
	V_{\HF}(\Delta,Q\ {\rm or}\ \theta)=\frac{1}{2\pi}\sum_{\rm
	bands} \left( \int_{E(k)<E_{\rm F}}E(k)dk \right)
	+\frac{\Delta^{2}}{4U}+\frac{1}{4}Un^{2}.
	\label{integral}
\end{equation}
For the \SSDW\ energies \eref{EkS} this is evaluated in terms of the
elliptic integral of the second kind, $E(\phi,k)$ (Gradshteyn and
Ryzhik 1980):
\begin{eqnarray}
\fl	\int_{0}^{\kF}E_{\rm SSDW}(k)dk =
	-2t\cos(Q/2)\sin\kF
	\nonumber \\
\fl	\pm \frac{\Delta}{2}\left[ \sqrt{1+p^{2}}\
	E\left(\tan^{-1}[\sqrt{1+p^{2}}\tan\kF],p/\sqrt{1+p^{2}}\right)-
	\frac{p^{2}\sin\kF\cos\kF}{\sqrt{1+p^{2}\sin^{2}\kF}} \right]
	\label{elliptic}
\end{eqnarray}
where
\begin{equation}
	p=(4t/\Delta)\sin(Q/2).
	\label{p}
\end{equation}
We evaluate the \DSDW\ energies by numerical integration of the band
structure \eref{EkD}.

For each point $(n,U)$ in the phase diagram, we minimize the energies
of the \SSDW\ and \DSDW\ states with respect to the parameters
$(\Delta,Q)$ and $(\Delta,\theta)$ respectively.  Because of perfect
nesting, a non-zero \SSDW\ solution always exists in some $Q$ interval
for $U>0$.  The \DSDW\ phase requires a little more care, as the
energy gain is small and a non-trivial solution is absent in much of
the phase diagram.

\section{Phase diagram \label{pd}}
\subsection{Energies}
\Fref{energy}(a) shows the \SSDW\ and \DSDW\ \HF\ energies for $U=4t$
and $0\leq n\leq 1$.  Because of particle-hole symmetry, the range
$1<n\leq 2$ contains no further information and will not be discussed
further; references to quarter filling also apply to three-quarter
filling.  For comparison, the top curve in the figure shows the energy
of the paramagnetic solution $(\Delta=0)$ and the bottom curve is
Shiba's (1972) numerical solution of the Lieb-Wu integral equations
for the exact ground state energy of the one-band one-dimensional
Hubbard model.  While the \HF\ energies are not a good approximation
to the exact ground state of this model, their relevance is discussed
in section \ref{stab}.

The \SSDW\ ground state corresponds to saturated ferromagnetism for
small $n$.  The pitch $Q$ then increases continuously
from 0, reaching $\pi$ at half filling $(n=1)$. For small $U$, the nesting
condition implies $Q\approx 2\kF=\pi n$; as $U$ increases, the \SSDW\
moves to smaller $Q$, reducing double occupancy. The
\DSDW\ phase is stable with respect to the paramagnetic phase
in a small range of fillings, and is only stable with respect to
\SSDW\ in a very narrow region near quarter filling $(n=1/2)$.  The
\DSDW\ is most stable at exactly quarter filling, when the energy is
minimized at the collinear \DSDW\ $\theta=0$.  The angle $\theta$
increases smoothly and monotonically (initially linearly) with
deviation from quarter filling.

The above discussion applies for the range $0<U<U_{1}\approx 5.58t$;
for $U>U_{1}$ the \DSDW\ state always has higher energy than the
competing \SSDW\ ground state (with $Q<\pi/2$).  Only for $U\approx
4t$ can the \DSDW\ region be seen in such a plot.  It is difficult to
find a \DSDW\ solution numerically for small $U$.  A fit in the range
$t\leq U \leq 3t$ suggests that the energy gain in forming a \DSDW\
has the form $\exp(-U_{2}/U)$, with $U_{2}\approx 17t$.

Suzumura and Tanemura (1995) have recently reported results on the
\HF\ ground state and excitations of the quarter-filled Hubbard model.
They consider a different class of configurations, namely collinear
$Q=\pi/2$ spin density waves.  Such states are pinned to the lattice
by a small commensurability energy dependent on the phase of the wave.
Their ground state, with phase $\pi/4$, is precisely the \DSDW\ (of
energy $-0.681t$ at $U=4t$, as calculated here).  At all other phase
angles, a $Q=\pi$ charge density wave coexists with the spin density
wave.  Maximum energy ($-0.679t$) occurs at zero phase, corresponding
to a magnetic configuration $\uparrow\cdot\downarrow\cdot\ldots$.
This is still less than our minimum \SSDW\ energy of $-0.674t$.  They
find the value of $U_{2}/t$ to be $4\pi\sqrt{2}\approx 17.8$, compared
with our numerical fit of 17.

\subsection{Phase diagram}
\Fref{energy}(b) shows the phase diagram in the $(n,U)$ plane.  For
$U>U_{\rm c}(n)$, indicated by the bold line, the ground state is
saturated (strong) ferromagnetism (\FM) with the upper band empty.
The absence of unsaturated ferromagnetism is a consequence of the
band-edge divergence of the one-dimensional density of states.  The
energy of this state (for $n\leq 1$) is independent of $U$, as there
is no double occupancy:
\begin{equation}
	E_{\rm HF}=-\frac{2}{\pi}t\sin n\pi.
	\label{EFM}
\end{equation}
The phase boundary (which appears to be linear for small $U$), can be
calculated by expanding the band structure \eref{EkS} to $O(Q^{2})$
and integrating; the line where $\partial^{2}V_{\rm
HF}(\Delta,Q)/\partial Q^{2}$ vanishes is
\begin{equation}
	U_{\rm c}(n)=\frac{2\pi n-\sin 2\pi n}{n \sin n\pi}.
	\label{Uc}
\end{equation}
The limit $n\rightarrow 1$ gives $U_{\rm c}\rightarrow \infty$.  Thus,
in the infinite-repulsion limit, \FM\ is (as expected) stable for all
$n\ne 1$.  At half filling $(n=1)$ the stable state is \AFM.
\Fref{EUn} shows the energy of the uniform ground state (\FM, \SSDW\
or \DSDW) as a function of $n$ for various values of $U$.  Where the
uniform ground state is \FM, the energy coincides with the $U=\infty$
curve \eref{EFM}.  \DSDW\ is, as already mentioned, stable in only a
small region of the phase diagram.  All states are metallic apart from
the \AFM\ at half filling and the \DSDW\ at quarter filling, where the
Fermi level lies in a gap.

\subsection{Phase separation\label{ps}}
A curious feature of the \SSDW\ energy is that it is not a convex
function of $n$ near $n=1$.  The uniform phase is therefore {\it
unstable} to phase separation.  Such an instability in the Hubbard
model was proposed by Visscher (1974) and is analysed here by means of
a Maxwell construction (Marder \etal 1990, Arrigoni and Strinati
1991).  The broken line $E_{\rm ps}(n)$ in \fref{energy}(a) coincides
with $E_{\rm HF}(n)$ at $n=1$ and is tangent to the curve at
$n=n_{1}(U)$, which is the boundary for phase separation.  For
$n_{1}<n<1$ the uniform \SSDW\ is unstable towards a state with volume
fraction $(n-n_{1})/(1-n_{1})$ of the $n=1$ \AFM\ phase and
$(1-n)/(1-n_{1})$ of the $n=n_{1}$ \SSDW\ phase.  The energy of the
phase-separated state is
\begin{equation}
	E_{\rm ps}(n)=\frac{(n-n_{1})E_{\rm HF}(1)
	   +(1-n)E_{\rm HF}(n_{1})}{1-n_{1}}
	<E_{\rm HF}(n).
	\label{Eps}
\end{equation}
The boundary is shown as a broken line superposed on the phase
diagram in \fref{energy}(b). To the right of the
boundary the uniform phase is unstable towards phase separation
between the \AFM\ phase and the \FM\ or \SSDW\ phase at the boundary.
For $U>8t$ (approximately), the only phases are \FM\ and \AFM.
Andriotis \etal (1993), in supercell calculations on the
one-dimensional Hubbard model with collinear moments, see a phase
separation of just this form for large $U$.  In one dimension the
phase separation is an artefact of the \HF\ approximation, as the
exact energy is a convex function of $n$ (Shiba 1972).  If phase
separation does occur in the Hubbard model in higher dimensions, it is
a consequence of purely short-range interaction and would be
suppressed by long-range Coulomb repulsion.  On the other hand, we do
not know whether the domains are of macroscopic size.  The instability
might signal phase separation at a more local level --- the migration
of holes to antiferromagnetic domain walls in a soliton lattice.  This
possibility depends on the sign of the domain wall energy.

Because of the kink in the energy at half filling there is no
separation into $n>1$ and $n<1$ phases. A similar phase separation
does occur between the \SSDW\ and \DSDW\ phases,
implying that the pure \DSDW\ phase is only stable at exactly quarter
filling. To avoid complicating the diagram, this is omitted from
\fref{energy} but is discussed in section \ref{stab}.

\section{Discussion \label{disc}}
\subsection{Interpretation of the DSDW}
The \DSDW\ phase illustrates a number of physical phenomena. Firstly, it
is an extreme case of a commensurate soliton lattice: an
antiferromagnet with (uncharged) domain walls on alternate bonds.
Secondly, the collinear \DSDW\ is clearly stabilized near
quarter filling by the opening of a gap at the Fermi energy (\fref{bands}).

The \HF\ energy can be expanded in the field:
\begin{equation}
	V_{\rm HF}(\Delta, \{{\bf e}_{i}\})=V_{\rm HF}(0, \{{\bf e}_{i}\})
	+\frac{1}{N_{\rm a}}\Delta^{2}\sum_{ik}J_{k}{\bf e}_{i}\cdot{\bf
	e}_{i+k}+O(\Delta^{4}).
	\label{expand}
\end{equation}
The qualitative features of the phase diagram follow from a theorem
concerning the dependence of susceptibilities on band filling (Heine
and Samson 1980, 1983).  Consider the moment expansion of the Green
function
\begin{equation}
	G_{ij}(z)=[(z-H)^{-1}]_{ij}=\delta_{ij}z^{-1}+H_{ij}z^{-2}+
	\sum_{k}H_{ik}H_{kj}z^{-3}+\ldots,
	\label{moment}
\end{equation}
where $H$ is a tight-binding Hamiltonian.  The energy difference
between two configurations A and B changes sign $r$ times as a
function of band filling if the leading term in the difference of
Green functions has the asymptotic form
\begin{equation}
	{\rm Tr}\sum_{i}[G_{ii}^{\rm (A)}(z)-G_{ii}^{\rm (B)}(z)] \sim z^{-r-3}
	\label{asymp}
\end{equation}
for large $z$, where Tr is a trace over spin.  A consequence is that
the effective nearest-neighbour exchange interaction $J_{1}$ changes
sign twice in the band $0<n<2$, while $J_{2}$ changes sign four times.
At quarter filling, $J_{1}$ is small and ferromagnetic and $J_{2}$ is
larger and antiferromagnetic.  Longer-range interactions are smaller.
Hence the \DSDW\ region of the phase diagram of the Hubbard model
corresponds to the dimerized region of the Majumdar-Ghosh model
discussed in the introduction.

The same theorem, applied to nonlinear susceptibilities, describes the
state selection between collinear and noncollinear \DSDW. The leading
term in the $\theta$-dependence of the Green function must involve a
biquadratic term in the angles.  Leaving out the terms irrelevant to
the present argument and summing equivalent paths gives
\begin{eqnarray}
	{\rm Tr}G_{00}^{\theta}(z) & = & \ldots-\frac{1}{4}{\rm
	Tr}\left(\bDelta_{0}\cdot{\bf\sigma}t_{01}\bDelta_{1}\cdot{\bf\sigma}t_{10}
	\bDelta_{0}\cdot{\bf\sigma}t_{01}\bDelta_{1}\cdot{\bf\sigma}t_{10}\right)
	z^{-9}+\ldots \nonumber \\ & = &
	\ldots-\frac{1}{2}\Delta^{4}t^{4}[({\bf e}_{0}\cdot{\bf
	e}_{1})^{2}-|{\bf e}_{0}\times{\bf e}_{1}|^{2}]z^{-9}+\ldots
	\nonumber \\ & = &
	\ldots-\frac{1}{2}\Delta^{4}t^{4}\cos(2\theta)z^{-9}+\ldots.
	\label{GDSDW}
\end{eqnarray}
The energy difference between collinear and noncollinear \DSDW\ states
with equal $\Delta$ therefore has {\it six} zeros as a function of
band filling (and a $\cos 2\theta$ dependence).  This agrees with the
numerical results: the collinear phase is favoured near quarter, half
and three-quarter filling, and the non-collinear phase at other
values.  Similar behaviour is indeed seen in calculations on
$\gamma$-Mn (Long and Yeung 1986, 1987).  The two-dimensional analogue
seems to be the windmill configuration observed by Ichimura \etal
(1992) in a quarter filled square lattice.  We note that the
conclusion about the sign of the energy difference is not restricted
to small $\Delta$, where a Landau expansion of the form \eref{expand}
can be truncated at fourth order (Schulz 1990).

\subsection{Stability and relevance of solutions \label {stab}}
\begin{figure}
	\caption{Energy of the uniform HF ground state as function of
	band filling $n$ for various values of $U$.  The $U=\infty$ curve
	corresponds to FM.}
    \label{EUn}
\end{figure}
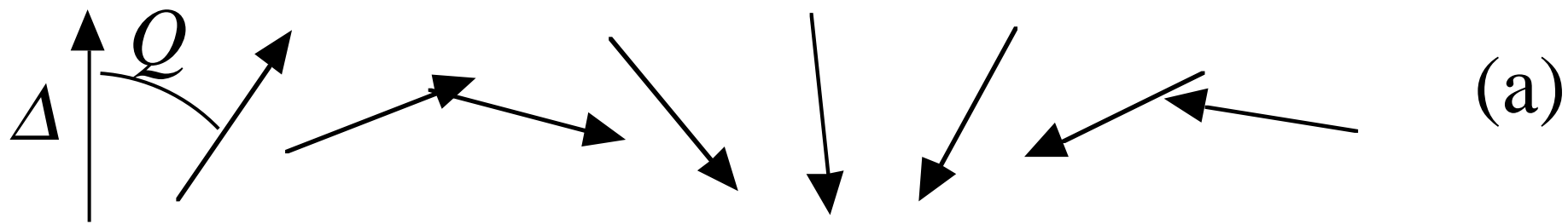
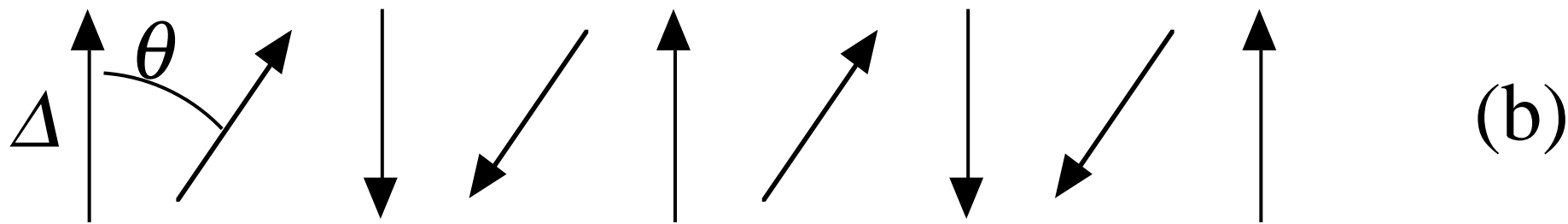
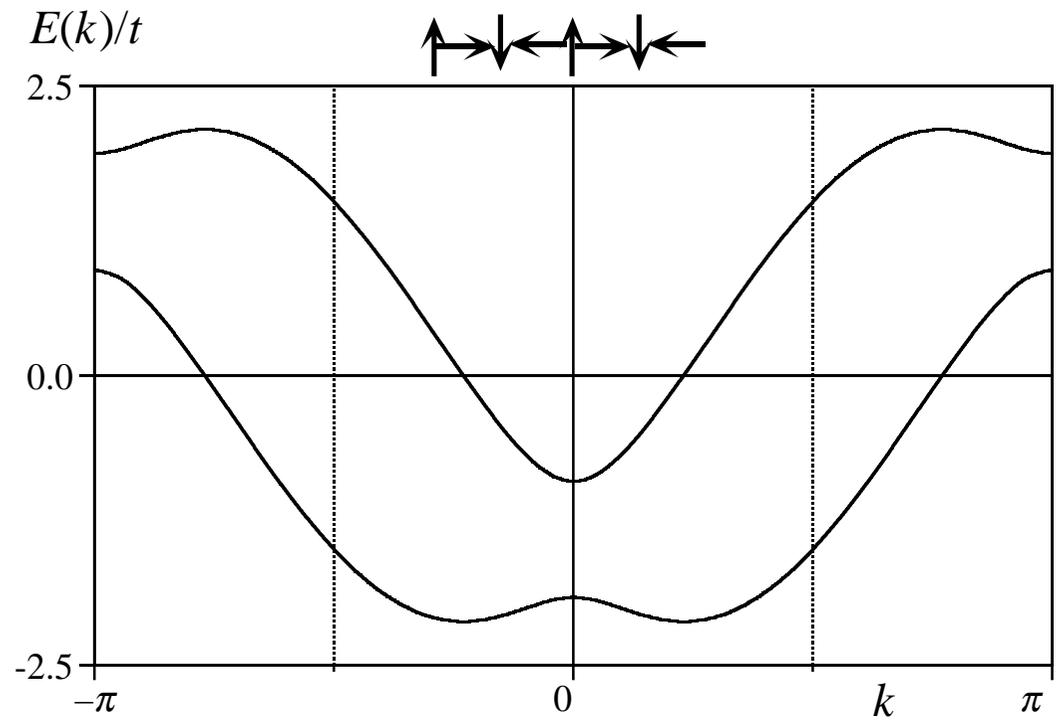
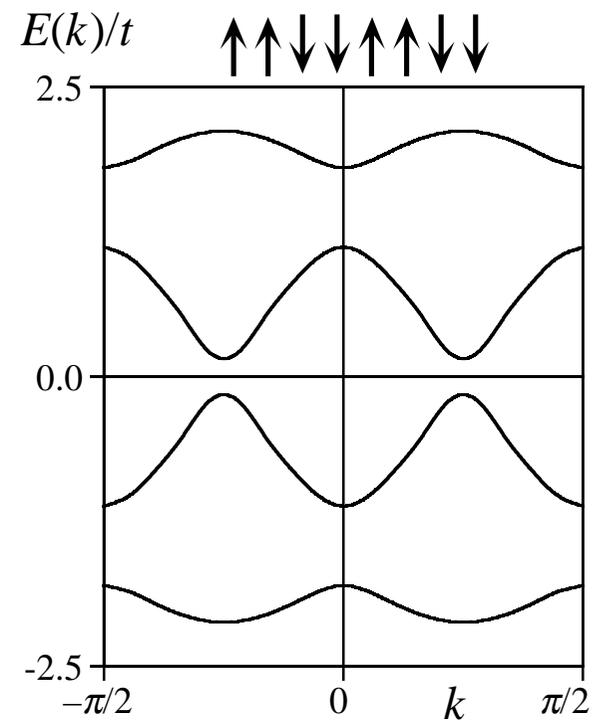
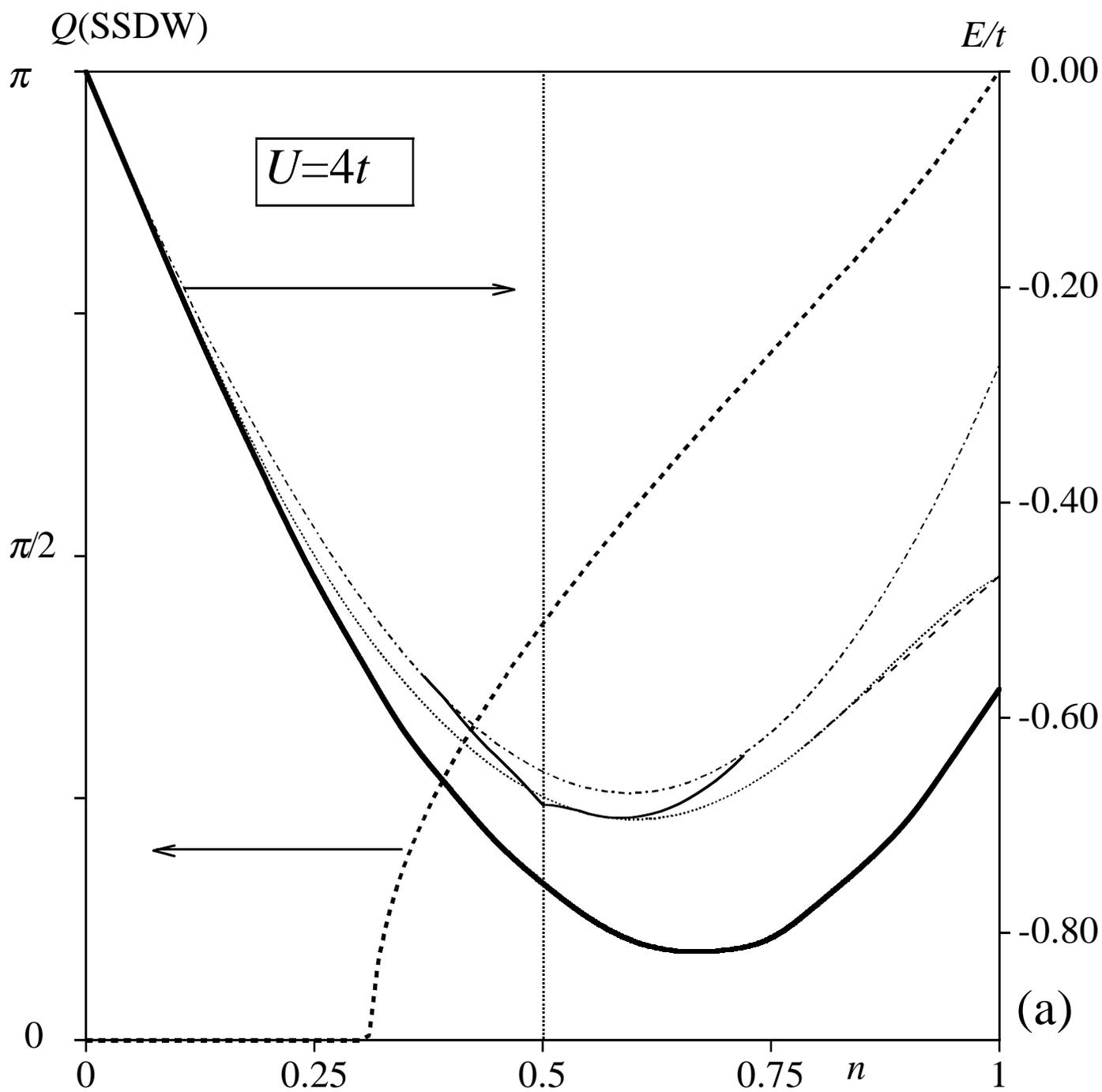
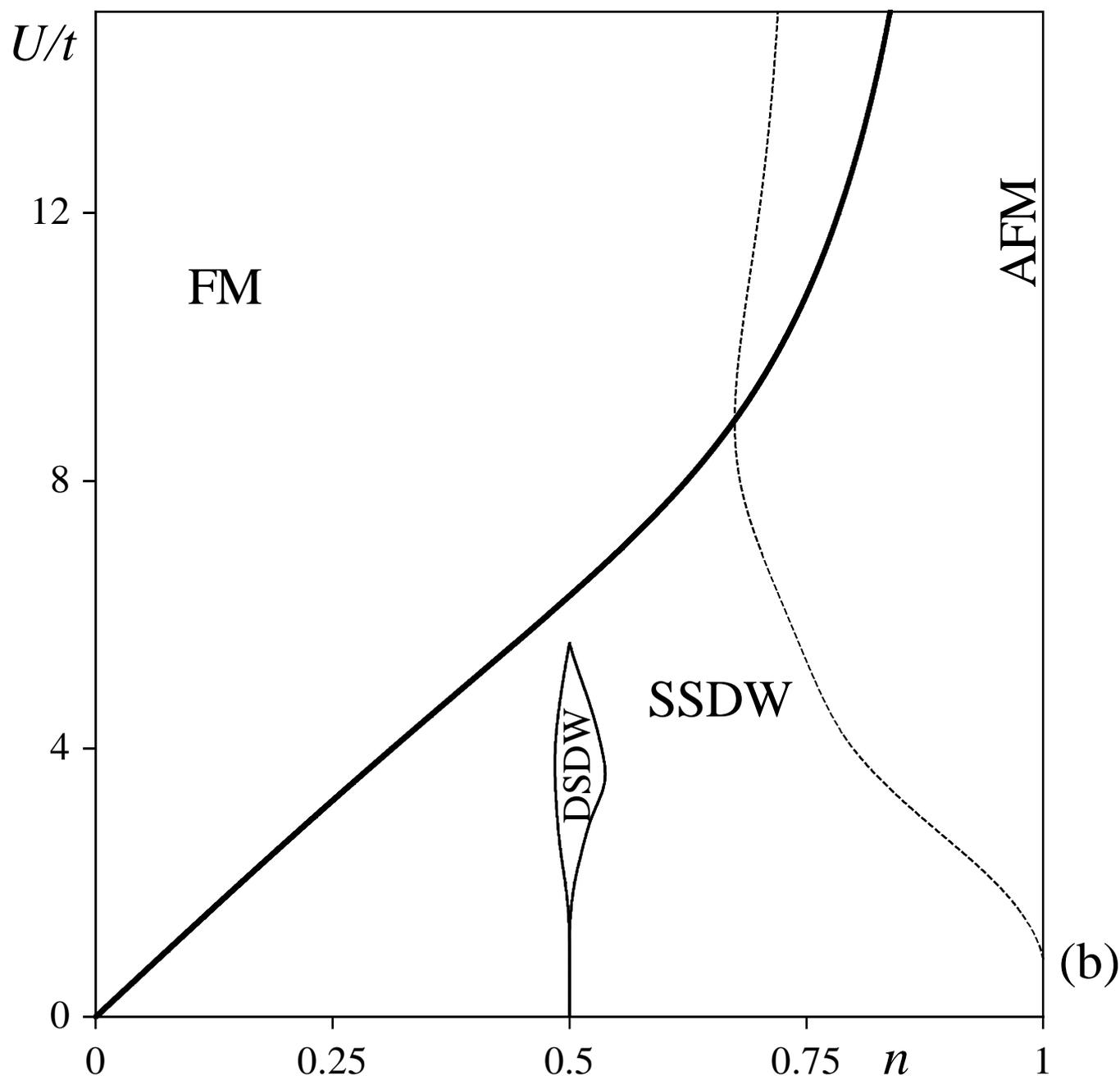

The solutions presented here (except at quarter and half filling) are
metallic.  In a one-dimensional system we would still expect an
instability towards a non-uniform state that opens a gap at the Fermi
energy.  Auerbach and Larson (1991) find just such an instability
towards a local increase in the spiral pitch in the $t-J$ model.  A
similar instability is seen in the two-dimensional Hubbard model (Zhou
and Schulz 1995).  Phase separation is a further indication of this
instability.  If the filling is rational, the distortion would be
commensurate; if it is irrational, an incommensurate distortion
leading to a Cantor set spectrum is a possible outcome (Ostlund and
Pandit 1984).  Early studies of such spectra were more concerned with
charge density waves, which, unlike the uniform \SSDW, modulate the
lattice potential.  A similar modulation in the latter case requires a
further distortion of the \SSDW, leading to coexisting charge-density
waves.  This will result in a small kink in the $E(n)$ curves as in
\fref{EUn}.  The true \HF\ energy, according to the Maxwell
construction, will be the convex hull of $E(n)$.  The ground-state
$Q$, plotted as a function of chemical potential for constant $U$,
would then be a staircase.  The \DSDW\ results and the small lock-in
energies suggest that any energy gain will be small, and the tongues
in the phase diagram will be correspondingly narrow.  However,
discussion of the fractal properties of this phase diagram will take
us rather far from the physics of the Hubbard model.

The results reported here are consistent with known exact results for
unrestricted generalized \HF\ (Bach \etal 1994).  These authors discuss
{\it inter alia} the \HF\ ground state of the repulsive Hubbard model
on a bipartite lattice.  This is \AFM\ for half filling, and \FM\ in
the $U\rightarrow\infty$ limit for all other fillings.  They cannot obtain
similar results in the interior of the phase diagram.  However, the
large gap and the work of Suzumura and Tanemura (1995) strongly
suggest that the collinear \DSDW\ is the true unrestricted \HF\ ground
state at quarter filling.

It must be admitted again that \HF\ calculations fail to give the
correct ground state symmetry, and give poor ground state energies;
quantum fluctuations destroy \AFM\ and \SSDW\ order in one dimension,
but may retain short-range correlations of that form.  The
higher-order correlations that distinguish the \DSDW\ would be harder
to see.  Agreement appears to be better in two dimensions where \HF\
solutions are a useful leading approximation (Mehlig 1993, Mehlig and
Fulde 1994).  For two and more dimensions the space to explore, even
for uniform configurations, is much larger and an exhaustive study
would be more time-consuming.  On the other hand, symmetry breaking is
possible in the ground state, so that \HF\ calculations may give a
ground state of the correct symmetry.  The paramagnetic phase will be
stable for small $U$ (except for the half-filled bipartite lattice),
and the unrestricted \HF\ energy $E(n)$ will be a smoother function
than in one dimension, as it will be more difficult to open a gap.
The calculations here can then be considered as a toy model,
nevertheless giving physically relevant predictions of spiral phases
and a special phase at quarter filling.

While the \HF\ results give little information on the integrable
one-band one-dimensional model, the approximation is more useful in
the degenerate case (which is not integrable).  Many-electron atoms
may be modelled by an $N$-band Hubbard Hamiltonian with inter-orbital
Coulomb repulsion $(U/4N)n_{i}^{2}$ and Hund's rule term
$-(I/N)S_{i}^{2}$, dependent on the total charge and spin on the atom.
There is no reason that these should be governed by the same coupling
constant for $N>1$.  In the case $U\gg I$, Coulomb repulsion will
suppress the phase-separated and non-uniform configurations, thereby
stabilizing the \SSDW\ against phase separation.  The \HF\ results for
uniform states are independent of $N$, and are the leading
approximation in the limit $N\rightarrow\infty$.

The energies obtained here may also be useful for finite temperature
properties.  In a study of the thermodynamics, Samson (1989) fitted
the \SSDW\ energies to an extended spherical model, obtaining
indications that the correlation functions fall more slowly with
temperature than in the Heisenberg model.  That approximation however
cannot distinguish the \DSDW\ from the $Q=\pi/2$ \SSDW; the small
energy gain suggests they would have little thermodynamic
significance.

While \HF\ solutions of the one-dimensional Hubbard model clearly
cannot give definitive answers to questions about higher dimensions,
they do provide a model in which many of the physical processes
underlying state selection operate in a transparent way.

\ack
The author is grateful to A S Alexandrov and M W Long for many helpful
comments on the manuscript.

\References
\item[] Andriotis A N, Economou E N and Soukoulis C M 1993 \JPCM {\bf
5} 4505--18
\item[] Arrigoni E and Strinati G C 1991 \PR B {\bf 44} 7455--65
\item[] Auerbach A and Larson B E 1991 \PR B {\bf 43} 7800--9
\item[] Bach V, Lieb E H and Solovej J P 1994 {\it J. Statist. Phys} {\bf
76} 3--89
\item[] Chubukov A V and Musaelian K A 1995 \PR B {\bf 51} 12605--17
\item[] Crockford D J, Bird D M and Long M W 1991 \JPCM {\bf 3}
8665--82
\item[] Dzierzawa M 1992 \ZP B {\bf 56} 49--52
\item[] Eriksson A B, Einarsson T and \"Ostlund S 1995 \PR B {\bf 52}
3662--75
\item[] Fujita M, Ichimura M and Nakao K 1991 \JPSJ {\bf 60} 2831--4
\item[] Gradshteyn I S and Ryzhik I M 1980 {\it Table of Integrals,
Series, and Products} (London: Academic) formula 2.597.2
\item[] Haldane F D M 1982 \PR B {\bf 25} 4925--8
\item[] Heine V and Samson J H 1980 \JPF {\bf 10} 2609--16
\item[] \dash 1983 \JPF {\bf 13} 2155--68
\item[] Henley C L 1987 \JAP {\bf 61} 3962--4
\item[] \dash 1989 \PRL {\bf 62} 2056--9
\item[] Hirai K and Jo T 1985 \JPSJ {\bf 54} 3567--70
\item[] Hu F, Sarker S K and Jayaprakash C 1994 \PR B {\bf 50}
17901--9
\item[] Ichimura M, Fujita M and Nakao K 1992 \JPSJ {\bf 61} 2027--39
\item[] Korenman V, Murray J L and Prange R E 1977 \PR B {\bf 16}
4032--47
\item[] Lieb E H and Wu F Y 1968 \PRL {\bf 20} 1445--8
\item[] Long M W 1989 \JPCM {\bf 1} 2857--74
\item[] Long M W and Yeung W 1986 \JPF {\bf 16} 769--90
\item[] \dash 1987 \JPC {\bf 20} 5839--66
\item[] Machida K and Fujita M 1984 \PR B {\bf 30} 5284--99
\item[] Majumdar C K and Ghosh D K 1969 \JMP {\bf 10} 1388--98
\item[] Marder M, Papanicolaou N and Psaltakis G C 1990 \PR B {\bf 41}
6920--32
\item[] Mehlig B 1993 \PRL {\bf 70} 2048
\item[] Mehlig B and Fulde P 1994 \ZP B {\bf 94} 335--9
\item[] Ostlund S and Pandit R 1984 \PR B {\bf 29} 1394--414
\item[] Penn D R 1966 \PR {\bf 142} 350--65
\item[] Samson J H 1989 \JPCM {\bf 1} 6717--29
\item[] \dash 1995 \JMMM {\bf 140--144} 205--6
\item[] Schulz H J 1990 \PRL {\bf 64} 1445--8
\item[] Schulz H J, Ziman T A L and Poilblanc D 1994 in {\it Magnetic
Systems with Competing Interactions} ed H~T~Diep (Singapore: World Scientific)
\item[] Shender E F 1982 {\it Sov. Phys.-JETP} {\bf 56} 178--84 (translation of
{\it Zh.  Exsp.  Teor.  Fiz.} {\bf 83} 326--37)
\item[] Shiba H 1972 \PR B {\bf 6} 930--8
\item[] Small L M and Heine V 1984 \JPF {\bf 14} 3041--52
\item[] Suzumura Y and Tanemura N 1995 \JPSJ {\bf 64} 2298--301
\item[] Tonegawa T and Harada I 1987 \JPSJ {\bf 56} 2153--67
\item[] Verg\'{e}s J A, Louis E, Lomdahl P S, Guinea F and Bishop A R
1991 \PR B{\bf 43} 6099--108
\item[] Visscher P B 1974 \PR B {\bf 10} 943--5
\item[] Zeng C and Parkinson J B 1995 \PR B {\bf 51} 11609--15
\item[] Zhou C and Schulz H J 1995 \PR B {\bf 52} R11557--60
\endrefs

\Figures

\begin{figure}
	\caption{Magnetization directions in (a) the SSDW and (b) the
	DSDW phases.}
	\label{configs}
\end{figure}

\begin{figure}
	\caption{Band structures for (a) the SSDW with $\Delta=t$ and
	$Q=\pi/2$ (equivalent to the DSDW with $\Delta=t$ and
	$\theta=\pi/2$) and (b) the DSDW with $\Delta=t$ and
	$\theta=0$.}
	\label{bands}
\end{figure}

\begin{figure}
	\caption{(a) HF results for the 1D Hubbard model as a function of
	band filling $n$ with $U=4t$.  The left-hand axis shows the pitch
	$Q$ of the SSDW (bold dashed line).  The right-hand axis shows the
	energies.  From top to bottom these are paramagnetic phase
	$(\Delta=0)$ (dash-dot line); SSDW phase (dotted); DSDW phase
	(full line); exact solution (bold line).  The short dashed line on
	the right is the Maxwell construction.  (b) The HF phase diagram
	for uniform phases (full lines).  Between the dashed line and
	$n=1$ the uniform phase is unstable to phase separation.}
	\label{energy}
\end{figure}

\end{document}